\documentclass[aps,prl,superscriptaddress,amsfont,graphicx,nofootinbib,preprintnumbers]{revtex4}%
\UseRawInputEncoding
\usepackage{color,graphicx,epsfig}
\usepackage{ifpdf}
\usepackage{amsmath}
\usepackage{bm}
\usepackage[english]{babel}
\usepackage{graphicx}%
\usepackage{amsfonts}%
\usepackage{amssymb}
\usepackage{braket}
\usepackage{hyperref}
\usepackage{enumerate}
\usepackage{cancel}

\usepackage{color}
\usepackage{ulem}
\usepackage{array}
\usepackage{ulem,fancyvrb}
\usepackage{xcolor}

\usepackage{dcolumn}
\usepackage{bm}
\usepackage{slashed}

\begin{document}

\title{Probing ultra-light dark photon from inverse Compton-like scattering}

\author{Liangliang Su}
\email{liangliangsu@njnu.edu.cn}
\affiliation{Department of Physics and Institute of Theoretical Physics, Nanjing Normal University, Nanjing, 210023, China}
\affiliation{School of Physics, Yantai University, Yantai 264005, China}

\author{Lei Wu}
\email{leiwu@njnu.edu.cn}
\affiliation{Department of Physics and Institute of Theoretical Physics, Nanjing Normal University, Nanjing, 210023, China}
\author{Bin Zhu}
\email{zhubin@mail.nankai.edu.cn}
\affiliation{School of Physics, Yantai University, Yantai 264005, China}

\date{\today}

\begin{abstract}
Dark photon not only provides a portal linking dark sector particles and ordinary matter but also is a well-motivated dark matter candidate. We propose to detect the dark photon dark matter through the inverse Compton-like scattering process $p+\gamma^\prime \to p+\gamma$. Thanks to the ultra-high energy primary cosmic rays, we find that such a method is able to probe the dark photon mass from $10^{-2}$ eV down to $10^{-19}$ eV with the expected sensitivity of eROSITA $X$-ray telescope, which can extend the current lower limit of dark photon mass from Jupiter's magnetic fields experiment by about three orders of magnitude.
\end{abstract}
\pacs{Valid PACS appear here}
\maketitle


\section{Introduction}
A large amount of astrophysical and cosmological observations suggest the existence of dark matter (DM). Investigation of the properties of DM is a common interest of particle physics, astrophysics, and cosmology. Currently, the only clues it gives us are through the gravitational effects, while there is no clear signal of the DM observed in the laboratory experiments, such as direct detection, indirect detection, and collider. The nature of the DM is still mysterious.

Among various well-motivated DM candidates, the Weakly Interacting Massive Particles (WIMP) have been intensively studied. However, the null results from direct detection, indirect detection and collider experiments have produced the strong constraints on some popular WIMP models. This has motivated to study other DM possibilities in recent years. One of the simplest hypotheses for DMs is that it belongs to a ``hidden sector'' or ``dark sector'' secluded from the Standard Model (SM)~\cite{Fayet:1980ad,Fayet:1980rr,Okun:1982xi,Georgi:1983sy,Fayet:1990wx,Pospelov:2008zw}, whose interactions could be mediated by a dark photon in an additional Abelian gauge group $U_{D}(1)$. The dark photon couples with the ordinary photon through the kinetic mixing~\cite{Holdom:1985ag}, and thus provides a portal linking the SM and new physics, which enables the dark photon can be detected in various experiments ~\cite{An:2013yfc,Redondo:2013lna,Liu:2014cma,An:2014twa,Aguilar-Arevalo:2016zop,Bhoonah:2018gjb,She:2019skm,McDermott:2019lch,Flambaum:2019cih,Garcia:2020qrp,Caputo:2020bdy,Witte:2020rvb,An2020jmf,Fitzpatrick:2021cij}. In the past decades, great efforts have been devoted to hunting for the dark photon (see the recent review~\cite{Fabbrichesi:2020wbt,Caputo:2021eaa}).

Recently, a number of novel production mechanisms, including the misalignment mechanism~\cite{Nelson:2011sf,Arias:2012az}, the parametric resonance production~\cite{Dror:2018pdh}, the tachyonic instability~\cite{Agrawal:2018vin,Co:2018lka,Bastero-Gil:2018uel} and the decay of a cosmic string network~\cite{Long:2019lwl}, have been proposed to efficiently produce the dark photons with a sub-eV mass down to the fuzzy dark matter scale $\sim 10^{-21}$ eV. Detections of such light dark photon dark matter have been studied in high precision spectroscopy~\cite{Jaeckel:2010xx,Danilov:2018bks}, helioscope~\cite{Redondo:2008aa,Ehret:2010mh,Betz:2013dza}, gravitational wave observation~\cite{Guo:2019ker,Co2021rhi}. Besides, the ultralight dark photon DM can also affect early cosmological observables and the evolution of astrophysical objects. For example, the conversion of dark photon and photon leaves observable imprints on the CMB background. In particular, the heating of the intergalactic medium around the epoch of helium ($\mathrm{He}^{++}$) reionization and the depletion of dark matter broaden the region of dark photon dark matter mass~\cite{McDermott:2019lch}. However, these bounds are based on some certain assumptions of the universe history, which is challenged by the recent different measurements of Hubble constant in early and late universe~\cite{DiValentino:2021izs}. Likewise, the heating/cooling rate of cold Galactic Center gas clouds can also constrain dark photon dark matter by requiring the heating rate from dark matter not exceeding the radiative cooling rate of the gas cloud ~\cite{Bhoonah:2018gjb,Wadekar2019xnf}. However, this constraint depends on the number density of elements in the gas cloud, such as electron number density $n_e$. Different results of calculating this quantity exist in~\cite{Bhoonah:2018gjb} and~\cite{Wadekar2019xnf}.

In this work, we propose an alternative way to probe the ultra-light dark photon dark matter through the interaction of the dark photon with the proton in the primary cosmic rays (CRs). Since the coupling of the dark photon with the ordinary matter is small, we consider the inelastic scattering process, $p+\gamma^\prime \to p+\gamma$, instead of the elastic scattering process. The energy of photon in the final states is determined by the dark photon mass and the proton energy in the CRs. Note that the latter is negatively correlated with the former for a given photon energy. In addition, there exists the ``GZK cut-off'' of the energy spectrum of the CRs, which arises from the energy loss of CRs during its scattering with the CMB photon~\cite{Greisen:1966jv,Zatsepin:1966jv}. Due to such a cut-off, the $X$-ray telescope is the most suitable experiment to detect the ultra-light dark photon in the mass range we are interested in. Nowadays, several $X$-ray telescopes have been launched, such as Chandra~\cite{Weisskopf:2001uu}, XMM-Newton~\cite{Turner:2000jy}, and the extended ROentgen Survey with an Imaging Telescope Array (eROSITA).~\cite{2012arXiv1209.3114M}. In the soft $X$-ray band (0.2-2.3 keV), eROSITA has a better outstanding energy resolution and field of view. The eROSITA all-sky survey will have reached a sensitivity more than 25 times higher than ROSITA all-sky survey at the end of eROSITA. Therefore, we will show the expected sensitivity our proposal in the eROSITA experiment.

\section{Theoretical Framework}

\begin{figure}[ht]
  \centering
  \includegraphics[height=5cm,width=8cm]{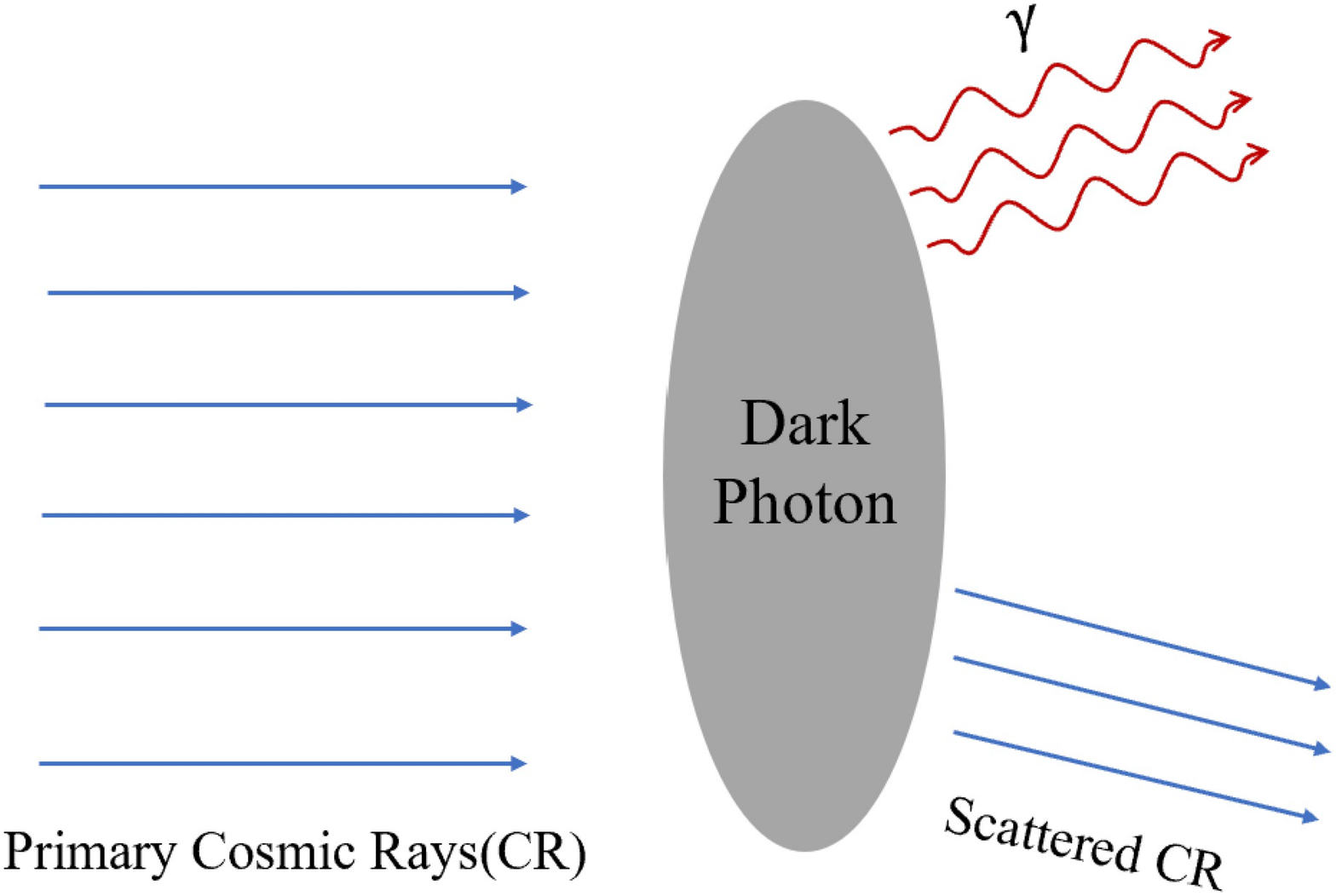}
\caption{Schematic diagram of the inverse Compton-like scattering process $\gamma^{\prime}(p_1) + p(p_2) \rightarrow \gamma(k_1) + p(k_2)$.}
\label{fig:scatter}
\end{figure}

The massive dark photon interacts with ordinary matter via the kinetic mixing with the ordinary photon. The relevant Lagrangian can be written as following, 
\begin{equation}
\begin{aligned}
 \mathcal{L} =& -\frac{1}{4} F_{\mu\nu} F^{\mu\nu}- e J_{\mu} A_{\mu} \\
 & -\frac{1}{4} F_{\mu\nu}^{\prime} F^{\prime\mu\nu} + \frac{1}{2}m_{\gamma^{\prime}}^2 A_{\mu}^{\prime} A^{\prime \mu} -  \epsilon  e J_{\mu}A^{\prime \mu}
\end{aligned}
\end{equation}
where $F_{\mu\nu}$ and $F_{\mu\nu}^{\prime}$ are the field strength tensors of the ordinary photon and the dark photon. The dark photon mass can originate from the Stueckelberg or the dark Higgs mechanisms. There are various constraints on the dark photon mass. Among them, the large-scale structure formation that requires the de Broglie wavelength of the dark photon does not exceed the DM halo size of the smallest dwarf galaxies produces a lower bound on the dark photon mass, $m_{\gamma^\prime} \gtrsim 10^{-22}$ eV~\cite{Hui:2016ltb}.  The last term represents the interaction of the dark photon $A^{\prime\mu}$ with the electromagnetic current of the SM particles $J_{\mu}$ with the coupling $e\epsilon$.  This makes the inelastic scattering process of the primary CRs with the dark photon possible, $\gamma^{\prime}(p_1) + p(p_2) \rightarrow \gamma(k_1) + p(k_2)$, i.e., the inverse Compton-like scattering process, as shown in Fig.~\ref{fig:scatter}.

In the rest frame of the dark photon, the maximum and minimum energies of outgoing photon for a given energy of the proton $E_p$ in CRs are given by
\begin{equation}
    \omega_{\mathrm{min}/\mathrm{max}} = \frac{m_{\gamma^{\prime}}^2+2 m_{\gamma^{\prime}}E_p}{2 E_p+2 m_{\gamma^{\prime}}\pm 2 \sqrt{E_p^2-m_p^2}} \label{eq:omega}
\end{equation}
where $m_{\gamma^{\prime}}$ and $m_p$ are the mass of the dark photon and the energy of the proton, respectively. For the ultra-high energy proton, we can find that the maximum energy of photon radiation is close to the incoming energy of the proton, while the minimum energy of the photon emission is always larger than the half of the dark photon mass, $\omega \gtrsim m_{\gamma^{\prime}}/2$. 

The differential cross section $\mathrm{d}\sigma/\mathrm{d}\omega$ of this process is also calculated by
\begin{equation}
\begin{aligned}
\frac{\mathrm{d}\sigma}{\mathrm{d}\omega} =& \frac{1}{32\pi m_{\gamma^{\prime}}|\vec{p}_2|^2}\left(\frac{1}{6}\sum |\mathcal{M}|^2\right)\\
=&\frac{e^4 \epsilon^2}{6\pi m_{\gamma^{\prime}}^3 |\vec{p}_2|^2(2E_p+m_{\gamma^{\prime}})^2(2E_p+m_{\gamma^{\prime}}-2\omega)}\\
&\times [ m_{\gamma^{\prime}}^2(2E_p+m_{\gamma^{\prime}})^2(2E_p^2+2E_p m_{\gamma^{\prime}} +m_p^2+m_{\gamma^{\prime}}^2)\\
& + \omega^2(m_{\gamma^{\prime}}^2(2E_p+m_{\gamma^{\prime}})(6E_p+5m_{\gamma^{\prime}})+4m_p^4 +2m_p^2 m_{\gamma^{\prime}}(4 E_p+3m_{\gamma^{\prime}})) \\ 
& -2m_{\gamma^{\prime}}^2 \omega^3(2E_p+m_{\gamma^{\prime}}) \\ 
& -2m_{\gamma^{\prime}} \omega(2E_p+m_{\gamma^{\prime}})(4 E_p^2 m_{\gamma^{\prime}}+E_p(2 m_p^2+5m_{\gamma^{\prime}}^2) +2m_{\gamma^{\prime}}(m_p^2+m_{\gamma^{\prime}}^2))],
\end{aligned}
\label{eq:cross}
\end{equation}
where the momentum of CR proton follow the relation $|\vec{p}_2|^2 = E_p^2-m_p^2$ and $\left(\frac{1}{6}\sum |\mathcal{M}|^2\right)$ is the squared amplitude of the inverse Compton-like scattering process. It should be noted that the squared matrix element of the dark photon in Eq.~\ref{eq:cross} is different from that of the axion. Using the polarization sum for the massive vector boson, we can calculate the squared matrix element of the dark photon and expand it in the small mass of the dark photon, which is given by $ |M|^2 \sim 16 e^4 \epsilon^2 + \mathcal{O}(m_{\gamma^{\prime}})$. Similarly, for the axion, the squared matrix element will go like $|M|^2 \sim 8 \omega^2 m_a^2/(m_p^2-s)^2+\mathcal{O}(m_a^3)$. Then, in comparison with the axion case~\cite{Alonso-Alvarez:2019ssa,Dent:2020qev}, there will be an enhancement of $1/m_{\gamma^\prime}$ at the leading order in Eq.~\ref{eq:cross} when the dark photon mass is small. On the other hand, since the dark photon is a massive vector boson and its mass should be heavier than $10^{-22}$ eV, the Eq.~\ref{eq:cross} will not be infinitely large. We check and find that our cross section is the same as that of the Compton-like scattering process $e \gamma \to e \gamma^\prime$ up to a factor of $2/3$ because the dark photon in the initial states has three polarization states. Besides, we derive the unitarity bound for our process and find that the constraint is independent of the dark photon mass in the limit of $E_p \gg m_{\gamma^\prime}$ as shown in the appendix.

Next, we will calculate the flux of the outgoing photon in the Compton-like scattering process to derive the bounds on dark photon. The differential flux can be written as
\begin{equation}
\begin{aligned}
 \frac{\mathrm{d}\Phi}{\mathrm{d}\omega \mathrm{d} \Omega} =&  \frac{1}{4 \pi} \int_{l.o.s} \mathrm{d} l \int_{E_{p}^{\mathrm{min}}(\omega)} \mathrm{d} E_{p} \frac{\mathrm{d}^2 \Gamma_{p+\gamma^{\prime}\rightarrow p+\gamma}}{\mathrm{d}E_{p} \mathrm{d} \omega }\\
 =& \frac{D}{ m_{\gamma^{\prime}}} \int_{E_{p}^{\mathrm{min}}(\omega)} \mathrm{d} E_{p} \frac{\mathrm{d}\sigma}{\mathrm{d}\omega} \frac{\mathrm{d}\Phi_{p}}{\mathrm{d} E_{p} },
\end{aligned}
\label{eq:phi}
\end{equation}
where $\mathrm{d} \Phi_{p}/\mathrm{d}E_{p}$ is the number of particles per area/steradian/kinetic energy/time of the local interstellar spectrum. It should be noted that the minimum energy of the proton in CRs that is needed to produce a photon with the energy $\omega$ can be obtained from Eq.~(\ref{eq:omega}),
 \begin{equation}
 \begin{aligned}
 E_{p}^{\mathrm{min}} =& \frac{\omega-m_{\gamma^{\prime}}}{2}\\
 +&\frac{\sqrt{m_{\gamma^{\prime}} \omega^2(m_{\gamma^{\prime}}-2\omega)(m_{\gamma^{\prime}}^2-2 m_{\gamma^{\prime}} \omega -4 m_p^2)}}{2 m_{\gamma^{\prime}}(2 \omega-m_{\gamma^{\prime}})},
 \end{aligned}
 \end{equation}
which is approximately given as $E_{p}^{\mathrm{min}} \simeq \omega/2+m_p \sqrt{\omega/2m_{\gamma^{\prime}}}$ for the ultra-light dark photon dark matter mass. This implies that the minimal energy of the proton $E_{p}^{\mathrm{min}}$ in the CRs is proportional to $\sqrt{\omega}$ and inversely proportional to $\sqrt{m_{\gamma^\prime}}$ in the very low mass range. On the other hand, the primary energy spectrum of the CRs declines as the power law of the energy like $\sim E^{-2.7}$ when $E < \mathcal{O}(10^{15})$ eV~\cite{Gaisser:2013bla}. However, it will vanish at the ``GZK cut-off''. In other words, a lower threshold detector, such as $X$-ray telescope, is needed to search for the dark photon dark matter in a wide mass range. In principle, the electron in the CRs can also induce such a scattering process. However, its contribution is negligible for the very light dark photon because of the low ``cut-off'' energy of primary CR electron

In this work, we use the flux of the proton in the primary CRs in the four-component ``Global-Fit4'' model~\cite{Gaisser:2013bla,Xia:2020wcp},
\begin{equation}
    \frac{\mathrm{d}\Phi_p}{\mathrm{d}E_p} = \sum_{i=1}^{4} c_i E_p^{-\alpha_{i}} \exp\left(-\frac{E_p}{R_i}\right)
\label{crflux}
\end{equation}
where $c_i$, $\alpha_{i}$ and $R_i$ represent the normalization constants, the integral spectral indexes and the rigidity cut-off with the component index $i$. The values of these parameters are listed in the Table.~\ref{table:flux}.
\begin{table}[ht]
\begin{tabular}{c|cccc}
\hline
           \textbf{Global-Fit4} & \begin{tabular}[c]{@{}c@{}}$R_1$\\ (120 TeV)\end{tabular} & \begin{tabular}[c]{@{}c@{}}$R_2$ \\ (4 PeV)\end{tabular} & \begin{tabular}[c]{@{}c@{}}$R_3$\\  (1.5 EeV)\end{tabular} &
           \begin{tabular}[c]{@{}c@{}}$R_4$\\  (40 EeV)\end{tabular} \\ \hline
$c_i$      & 7000                                                      & 150                                                      & 12                                                       & 1.2  \\ \hline
$\alpha_i$ & 2.66                                                      & 2.4                                                      & 2.4                                                      & 2.4
\\ \hline
\textbf{H4a} & \begin{tabular}[c]{@{}c@{}}$R_1$\\ (4 PeV)\end{tabular} & \begin{tabular}[c]{@{}c@{}}$R_2$ \\ (30 PeV)\end{tabular} & \begin{tabular}[c]{@{}c@{}}$R_3$\\  (60 EeV)\end{tabular}  \\ \hline
$c_i$      & 7860                                                      & 20                                                      & 200                                                        \\ \hline
$\alpha_i$ & 2.66                                                      & 2.4                                                      & 2.6
\\ \hline
\end{tabular}
\caption{The normalization constants $c_i$, the integral spectral indexes $\alpha_i$ and the rigidity cut-off $R_i$ in the four-component ``Global-Fit4'' and "H4a" model of proton flux are taken as the Ref.~\cite{Gaisser:2013bla,Xia:2020wcp}.} 
\label{table:flux}
\end{table}

The $D$-factor in Eq.~\ref{eq:phi} that depends on the DM density distribution can be defined by the line-of-sight of integral of DM density of the Milky Way halo~\cite{Evans:2016xwx,Dekker:2021bos},
\begin{equation}
    D(\phi) = \int_{l.o.s} \mathrm{d} l \rho_{\mathrm{DM}}(r(l,R,\phi)).
\end{equation}
    where $l$ is the line-of-sight distance and the $r(l,R,\phi) = \sqrt{l^2+R^2-2 l R \cos{\phi}}$ is the radial distance from the Galactic Center (GC). $R$ is the distance from GC to the Earth and $\phi$ is the polar angle between the Earth-GC axis and line-of-sight direction. In this paper, we assume that the dark matter density $\rho_{\mathrm{DM}}$ satisfy the Navarro-Frenk-White (NFW) profile distribution\textcolor{blue}:$\rho_{\mathrm{DM}} = \rho_s/((r/r_s)(r/r_s+1)^2) $, the scale radial and characteristic density $\rho_s$ refer to the Ref.~\cite{Dekker:2021bos}. Since only the DM that distributes in the region of the radial distance from the Earth of 1 $\mathrm{kpc}$ is considered, we roughly assume the cosmic-ray flux is isotropic. However, it should be mentioned that the isotropy assumptions of the cosmic-ray flux may be broken down in the high energy (more protons are believed to come from the direction of Ursa Major.) 
The dedicated study of the anisotropy is beyond the scope of this paper. Under the assumption that CRs distribution is isotropical in the galaxy, the $D$-factor can be approximately proportional to the product of the effective distance $D_{\mathrm{eff}}$ and local dark matter density $\rho^{\mathrm{local}}_{\mathrm{DM}}$~\cite{Bringmann:2018cvk,Wang2019jtk,Su2020zny,Flambaum2020xxo}.

In order to derive the constraints on the dark photon dark matter from $X$-ray experiments, we can simply require the theoretical prediction of the $X$-ray flux not exceeding the measured Cosmic X-ray Background (CXB)~\cite{Dent:2020qev}. However, this approach suffers from large uncertainties and may lead to an over-conservative estimation because the Active Galactic Nuclei (AGN) as main source of the CXB background may be underestimated. In Refs.~\cite{Foster:2021ngm,Dekker:2021bos}, a more accurate way was proposed. It is supposed that the diffuse CXB follows a power-law relation $C \times \omega^{-1.42}$ with normalization constant $C = 8.44 $ cts/$\mathrm{cm}^{2}$/$\mathrm{s}$ /$\mathrm{sr}$/$\mathrm{keV}$ at $\omega =1$ keV. On the other hand, the detector background has a flat energy spectrum and distributes over the sky isotropically, which is estimated as 1151 counts $\mathrm{s}^{-1}$ $\mathrm{sr}^{-1}$ $\mathrm{keV}^{-1}$ in Ref.~\cite{Dekker:2021bos}. By simulating the all-sky $X$-ray signal over each spatial pixel and energy bin, one can obtain 95\% C.L. exclusion limit from the joint-likelihood analysis. With their derived results, we can   
impose the constraints on dark photon dark matter in the photon energy $2< \omega <10$ keV.

\section{Numerical Results And Discussions}

\begin{figure}[ht]
  \centering
  \includegraphics[height=9cm,width=12cm]{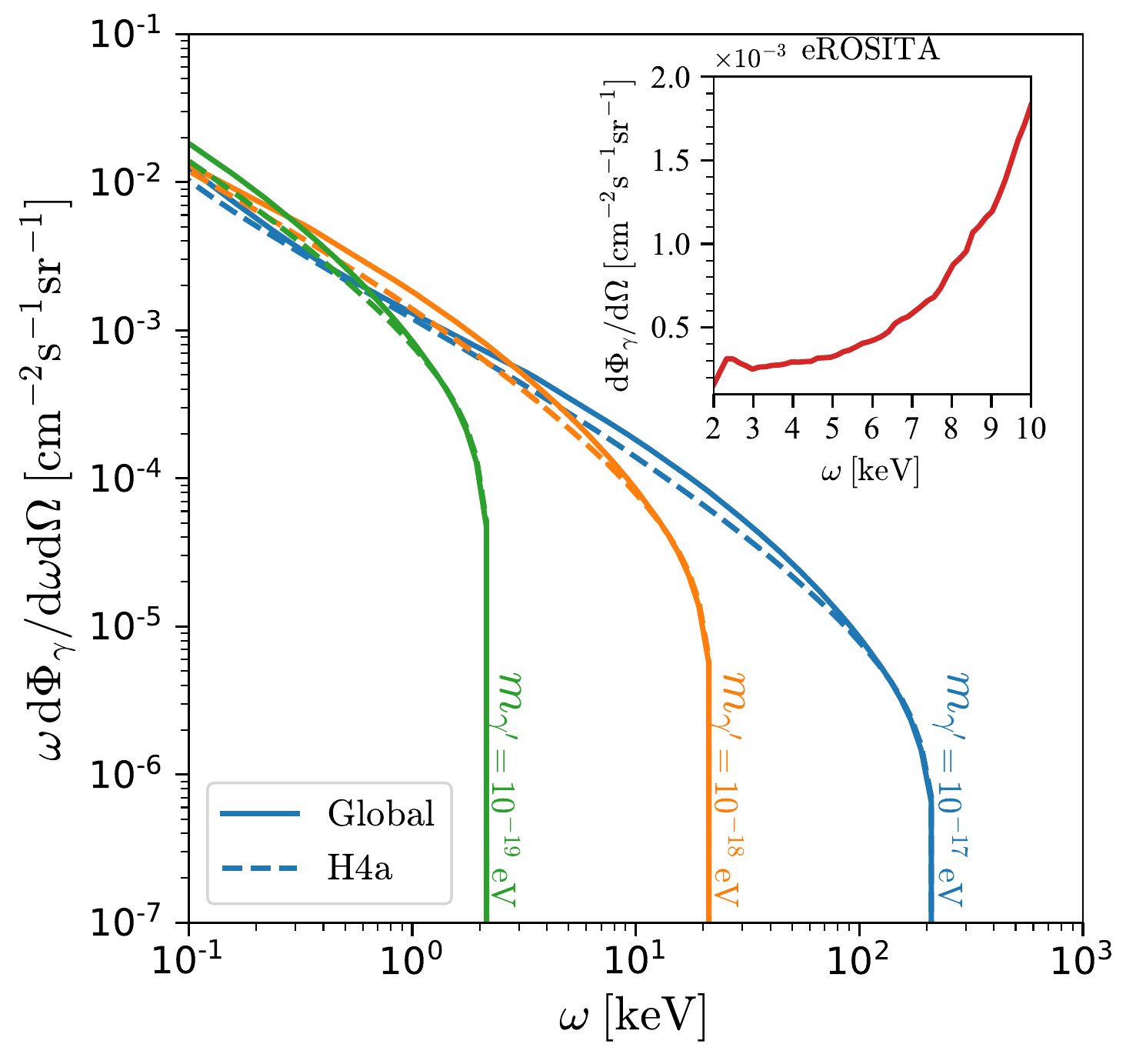}
\caption{The flux of the photon produced in the scattering process of $\gamma^\prime +p \to \gamma+ p$ for different dark photon mass $m = 10^{-19}$ eV(green), $m = 10^{-18}$ eV(orange) and $m = 10^{-17}$ eV(blue). We take $D = 10^{22} $ GeV/$\mathrm{cm}^{2}$ and $\epsilon = 0.1$ as an example. Two different parametric functions of the CRs: "Global-fit4"(solid lines) and "H4a"(dashed lines) are used. In the upper right corner of the figure, the limit on the photon flux $\mathrm{d} \Phi_{\gamma}/\mathrm{d} \mathrm{\Omega}$  by using the analysis of the eROSITA experiment~\cite{Dekker:2021bos} is given as well.}
\label{fig:flux}
\end{figure}
In Fig.~\ref{fig:flux}, we show that the photon flux for the dark photon mass $m_{\gamma^\prime} = 10^{-19},\;10^{-18},\;10^{-17}$ eV in the photon energy range $0.1<\omega<10^3$ keV. In the calculations, we choose $D = 10^{22} $ GeV/$\mathrm{cm}^{2}$ and the kinetic mixing parameter $\epsilon = 0.1$ as an example. The different values of these two parameters only change the magnitude of the photon flux, rather than distorting the shape. To show the effect the choice of model has on our results, we compare the results of two different parametric functions of the proton flux ``Global Fit4'' and ``H4a'' in Eq.~\ref{crflux}, whose parameters are given in Tab.~\ref{table:flux}. From Fig.~2. we find that the two proton fluxes will give the similar results and thus we choose ``Global Fit4'' in our following calculations. However, there remain large uncertainties in high-energy cosmic ray flux data that haven't been accounted for. In Fig.~\ref{fig:flux}, we can also see that the flux of the photon for a given dark photon mass decreases sharply with its energy, which is caused by the ``GZK cut-off'' of the CRs. The endpoints of each curve depend on the dark photon mass. For example, the photon flux vanishes at $\omega \sim 2$ keV and $\omega \sim 200$ keV for dark photon with $m_{\gamma^{\prime}} = 10^{-19}$ eV and  $m_{\gamma^{\prime}} = 10^{-17}$ eV, respectively. It can be understood that the maximal value of $\omega$ in Eq.~\ref{eq:omega} increases with the dark photon mass for a given proton energy, such as the ``GZK cut-off''. Thus, in order to access lighter dark photon through our inverse Compton-like scattering process, a lower energy threshold detector is required. In the upper right panel of Fig.~\ref{fig:flux},  we also display the limits on the photon flux by using the analysis of eROSITA experiment, which requires $1.6 \times 10^{-4} <\mathrm{d} \Phi_{\gamma}/\mathrm{d} \mathrm{\Omega}< 1.8 \times 10^{-3}$ cm$^{-2}$s$^{-1}$sr$^{-1}$ in the energy range of $2 \;\mathrm{keV} <\omega<10 \; \mathrm{keV}$~\cite{Dekker:2021bos}.

\begin{figure}[ht]
  \centering
  \includegraphics[height=9cm,width=12cm]{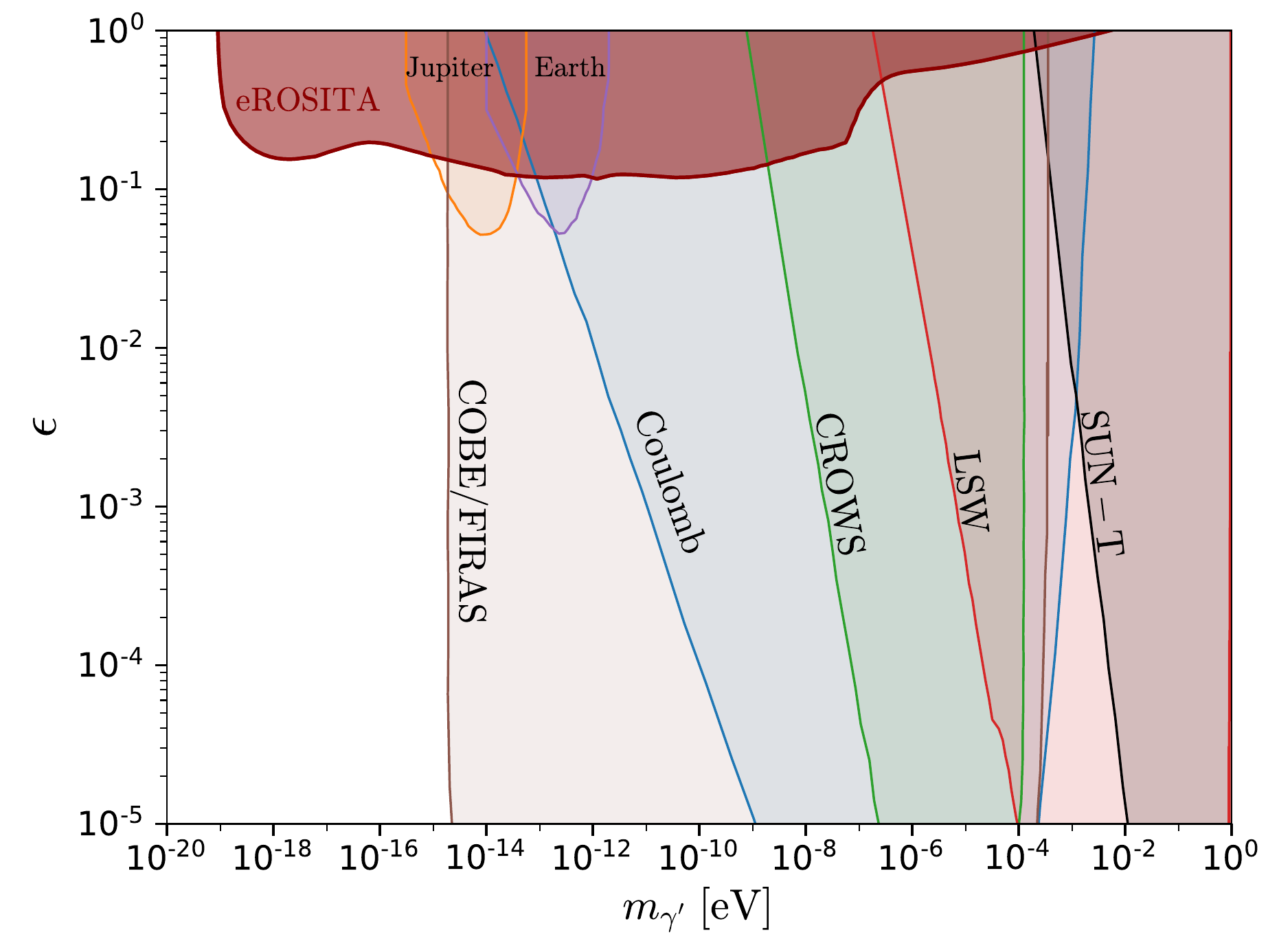}
\caption{The expected eROSITA limit by using inverse Compton-like scattering process $\gamma^{\prime}+p \to \gamma+p$ on dark photon dark matter mass $m_{\gamma^\prime}$ versus kinetic mixing parameter $\epsilon$. Other constraints are also shown, such as Jupiter (orange)~\cite{PhysRevLett.35.1402,Jaeckel:2010ni}, Earth (purple)~\cite{RevModPhys.43.277}, COBE/FIRAS (brown)~\cite{Caputo:2020bdy,Garcia:2020qrp}, Coulomb (blue)~\cite{Jaeckel:2010xx}, CROWS (green)~\cite{Betz:2013dza}, LSW (red)~\cite{Ehret:2010mh} and SUN-T (black)~\cite{An:2013yfc}.}
\label{fig:limit}
\end{figure}
In Fig.~\ref{fig:limit}, we present the bounds on the dark photon with the expected sensitivity of eROSITA $X$-ray telescope. In its energy range $2<\omega<10$ keV, the strongest bounds for ultra-light dark photon dark matter can be obtained at the photon energy $\omega = 2$ keV, as shown the upper right of Fig.~\ref{fig:flux}. Besides, we also plot other constraints on the dark photon dark matter with the mass less than 1 eV, including the magnetic fields of Jupiter~\cite{PhysRevLett.35.1402,Jaeckel:2010ni} and Earth~\cite{RevModPhys.43.277}, the CMB (COEB/FIRAS)~\cite{Caputo:2020bdy,Garcia:2020qrp}, the atomic experiments (Coulomb)~\cite{Jaeckel:2010xx}, the CROWS experiment~\cite{Betz:2013dza},  the light shining through a wall(LSW) experiments~\cite{Ehret:2010mh} and Solar lifetime (SUN-T)~\cite{An:2013yfc}. On the other hand, the constraints from the black hole superradiance are not shown because those constraints are only relevant in the limit of vanishing coupling. We can find that the dark photon mass from $10^{-6}$ eV to $10^{-19}$ eV with the kinetic mixing parameter $\epsilon \sim {\cal O}(0.1)$ can be excluded by the future eROSITA $X$-ray telescope experiment, which is lower than existing limit of dark photon mass from Jupiter's magnetic fields experiment by about three orders of magnitude. Besides, we also checked the constraint from the existing XMM-Newton data and fount it still much weaker than eROSITA. Finally it should be mentioned that the bound on the kinetic mixing parameter derived from our method is not as strong as others. However, it can cover a wide mass range in a single experiment and would be improved by future lower threshold telescopes. 

\section{Conclusions}
In this work, we investigated the potential of probing the ultra-light dark photon dark matter with $X$-ray produced from its inelastic scattering with the cosmic rays, $\gamma^{\prime}+p \to \gamma+p$. Due to the contribution of the ultra-high energy cosmic rays, we have found that the ultra-light dark photon dark matter mass can be excluded down to $10^{-19}$ eV by using the expected sensitivity of the eROSIRA $X$-ray telescope. Although the constraint of the kinetic mixing parameter is not stronger than others, it gives a new way to hunt for the ultra-light dark photon dark matter in a single experiment. In the future, with the continuous improvement of $X$-ray telescope experiments, our method will be able to further search for the lighter dark photon dark matter.

\section{Acknowledgments}
We thank Shan-Shan Weng for their helpful suggestions and discussions. This work is supported by the National Natural Science Foundation of China (NNSFC) under grant 
No. 11805161, by Jiangsu Specially Appointed Professor Program.

\section{Appendix}
In the section, we derive the unitarity bound on our model with the partial wave method. For the $2\rightarrow 2$ scattering process, we can have the partial wave $a_{fi}^J$, 
\begin{equation}
a_{f i}^{J}=\frac{1}{32 \pi} \int_{-1}^{1} \mathrm{~d} \cos \theta d_{\mu_{i} \mu_{f}}^{J}(\theta) \mathcal{T}_{f i}(\sqrt{s}, \cos \theta),
\end{equation}
where $J$ and $\sqrt{s}$ are the the total angular momentum and the center of mass energy (CM), respectively. The small Wigner $d$-function $d_{\mu_{i} \mu_{f}}^{J}$ is related with the polar scattering angle $\theta$ in the CM and the difference of helicities ($\mu_{i(f)}=\frac{1}{2}(\lambda_{i_1(f_1)}-\lambda_{i_2(f_2)})$) of the initial and final states. $ \mathcal{T}_{f i}$ is the scattering amplitude that is defined by $(2 \pi)^4 \delta^{(4)}(p_i-p_f) i \mathcal{T}_{f i}(\sqrt{s}, \cos \theta) = \langle f|S-1| i\rangle$. 
Especially, for $J=0$, $\mu_{i(f)} $ is zero. Thus, the partial wave can be written as 
\begin{equation}
a_{0}=\frac{1}{32 \pi} \int_{-1}^{1} \mathrm{~d} \cos \theta  \mathcal{T}_{f i}(\sqrt{s}, \cos \theta).
\end{equation}
Then, the unitarity condition of $S$ matrix, $S^{\dagger}S =1$, which implies 
\begin{equation}
    |a_0| < 1
\end{equation}

The key point is to calculate the scattering amplitudes $\mathcal{T}_{f i}$ between the initial and final state. For our scattering process $ p \bar{p} \to \gamma^\prime \gamma$, only the amplitudes $\mathcal{T}^{++++}$, $\mathcal{T}^{++--}$, $\mathcal{T}^{--++}$ and $\mathcal{T}^{----}$ contribute to the $J=0$ partial wave amplitudes. Thus, $a_{0}$ is rewritten as
\begin{equation}
a_{0}=\frac{1}{32 \pi} \int_{-1}^{1} \mathrm{~d} \cos \theta  \begin{pmatrix}  
  \mathcal{T}^{++++} & \mathcal{T}^{++--} \\  
  \mathcal{T}^{--++} & \mathcal{T}^{----}  
\end{pmatrix},
\end{equation}

\begin{figure}[ht!]
  \centering
  \includegraphics[height=6cm,width=8cm]{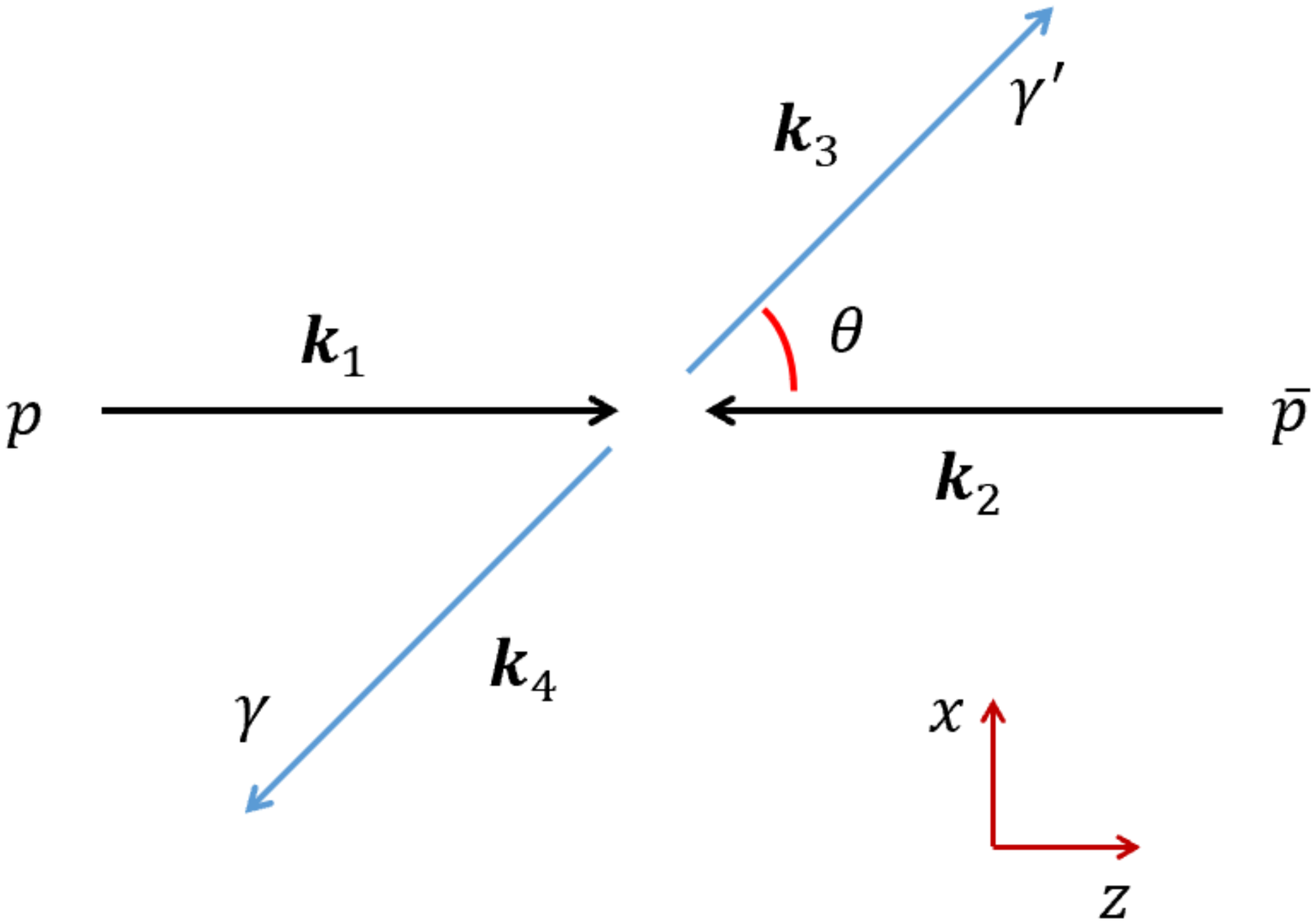}
\caption{The diagrammatic sketch of the $2\rightarrow 2$ scattering in the frame of the center of mass.}
\label{fig:scatter2}
\end{figure}

Next, we perform the calculations of the helicity amplitudes of the scattering process $p(k_1) \bar{p}(k_2) \to \gamma^\prime(k_3) \gamma(k_4)$, which is shown in Fig~\ref{fig:scatter2}.

In the CM, the four-momenta of the initial and final states are given by
\begin{equation}
    \begin{aligned}
    k_1 = & \frac{E_{cm}}{2}(1,0,0,\beta);\\
    k_2 = & \frac{E_{cm}}{2}(1,0,0,-\beta);\\
    k_3 = & (E_{cm} -\omega, \omega \sin{\theta},0, \omega \cos{\theta});\\
    k_4 = & (\omega, -\omega \sin{\theta},0, -\omega \cos{\theta}),
    \end{aligned}
\end{equation}
where $E_{cm}$ is the energy of the CM. $\omega$ and $\beta$ are the angular frequency of the photon and the speed of the proton, respectively. At the tree level, the amplitudes of $u$- and $t$-channels can be written as 
\begin{equation}
\begin{aligned}
i \mathcal{T}_{t}(\lambda_1,\lambda_2,\lambda_3,\lambda_4)&= - \epsilon e^2 \bar{v}(k_2,\lambda_2)\gamma_{\mu} \epsilon_{\gamma}^{\ast\mu}(k_4,\lambda_4) \\
&\times \frac{i(\cancel{k}_1\ -\cancel{k}_3 +m_p)}{(k_1-k_3)^2-m_p^2} \gamma_{\nu} \epsilon_{\gamma^{\prime}}^{\ast\nu}(k_3,\lambda_3) u(k_1,\lambda_1),
\end{aligned}
\end{equation}

\begin{equation}
\begin{aligned}
i \mathcal{T}_{u}(\lambda_1,\lambda_2,\lambda_3,\lambda_4) &= - \epsilon e^2 \bar{v}(k_2,\lambda_2)\gamma_{\mu} \epsilon_{\gamma^{\prime}}^{\ast\mu}(k_3,\lambda_3)\\
&\times \frac{i(\cancel{k}_1\ -\cancel{k}_4\ +m_p)}{(k_1-k_4)^2-m_p^2} \gamma_{\nu} \epsilon_{\gamma}^{\ast\nu}(k_4,\lambda_4) u(k_1,\lambda_1),
\end{aligned}
\end{equation}
where $\epsilon$ is the kinetic mixing parameter and $m_p$ is the mass of proton. $\gamma_{\mu(\nu)}$ is the Dirac matrix. By using the spinor $u(\bar{v})$ and the polarization vector $\epsilon_{\gamma^{\prime}}^{\ast\mu}(\epsilon_{\gamma}^{\ast\mu})$ in the Weyl representation,  we can obtain the helicity amplitudes as following
\begin{equation}
    \begin{aligned}
    \mathcal{T}^{++++} &=-\mathcal{T}^{----} \\
    &= \mathcal{T}_t(+,+,+,+)+\mathcal{T}_u(+,+,+,+)\\
   & = \frac{2 \epsilon e^2 m_p }{E_{cm}(E_{cm}^2-m_{\gamma^{\prime}}^2)(-1+\beta^2\cos^2{\theta})}\times (m_{\gamma^{\prime}}^2(2+\beta)\\
   &+2 E_{cm}^2(-1+\beta+\beta^2)+\beta(m_{\gamma^{\prime}}^2+2 E_{cm}^2\beta)\cos{2\theta})\\
    &\overset{\beta\rightarrow 1}{=} -\frac{2\epsilon e^2 m_p ( 2E_{cm}^2+ 3m_{\gamma^{\prime}}^2+(2E_{cm}^2+m_{\gamma^{\prime}}^2)\cos{2\theta}) \csc^2{\theta}}{E_{cm}(E_{cm}^2-m_{\gamma^{\prime}}^2)},
    \end{aligned}
\end{equation}

\begin{equation}
    \begin{aligned}
    \mathcal{T}^{++--} &= -\mathcal{T}^{--++} \\
    &=\mathcal{T}_t(+,+,-,-)+\mathcal{T}_u(+,+,-,-)\\
   & = \frac{2 \epsilon e^2 m_p }{E_{cm}(E_{cm}^2-m_{\gamma^{\prime}}^2)(-1+\beta^2\cos^2{\theta})}(m_{\gamma^{\prime}}^2(-2+\beta)\\
   &+2E_{cm}^2(1+\beta-\beta^2)+\beta(m_{\gamma^{\prime}}-E_{cm}^2\beta)\cos{2\theta})\\
    &\overset{\beta\rightarrow 1}{=} -\frac{4\epsilon e^2 m_p (2E_{cm}^2-m_{\gamma^{\prime}}^2) }{E_{cm}(E_{cm}^2-m_{\gamma^{\prime}}^2)},
    \end{aligned}
\end{equation}
where $m_{\gamma^{\prime}}$ is the mass of the dark photon. In the limit of $E_{cm} \gg m_{\gamma^{\prime}}$, the partial wave $a_{0}$ can be rewritten as
\begin{equation}
    a_{0} = \frac{\epsilon e^2 m_p}{4 \pi E_{cm}} \int_{-1}^{1} \mathrm{d} \cos{\theta} \begin{pmatrix}  
  -\cot^2{\theta} & -1 \\  
  1 &  \cot^2{\theta}
\end{pmatrix},
\end{equation}
in which the largest eigenvalue is approximately given by
\begin{equation}
    |a_0| \approx \frac{148 \epsilon \alpha m_p}{ E_{cm}}.
\end{equation}
Thus, we can derive the bound on the kinetic mixing parameter $\epsilon$,
\begin{equation}
    \epsilon < \frac{E_{cm}}{148 \alpha m_p},
\end{equation}
which shows that the unitarity will not be violated for a ultra-light dark photon. The similar results can also
be obtained in the scattering processes $p \gamma^\prime \to p\gamma$ and $p\bar{p} \rightarrow \gamma^{\prime} \gamma^{\prime}$. 

\bibliography{refs}

 \end{document}